\documentclass[sigconf]{acmart}
\usepackage{algpseudocode}
\usepackage{algorithm}

\AtBeginDocument{%
  \providecommand\BibTeX{{%
    \normalfont B\kern-0.5em{\scshape i\kern-0.25em b}\kern-0.8em\TeX}}}

\setcopyright{acmlicensed}
\copyrightyear{2025}
\acmYear{2025}
\acmDOI{XXXXXXX.XXXXXXX}

\acmConference[ArXiv PrePrint]{}{}{}
\acmISBN{978-1-4503-XXXX-X/18/06}

\begin{document}

\title{Intent Classification on Low-Resource Languages with Query Similarity Search}
\author{Arjun Bhalla}
\orcid{1234-5678-9012} 
\affiliation{%
  \institution{Bloomberg L.P.}
  \streetaddress{}
  \city{}
  \country{}}
\email{abhalla31@bloomberg.net}

\author{Qi Huang}
\orcid{1234-5678-9012} 
\affiliation{%
  \institution{Bloomberg L.P.}
  \streetaddress{}
  \city{}
  \country{}}
\email{qhuang127@bloomberg.net}

\renewcommand{\shortauthors}{Anonymous}

\begin{abstract}
  Intent classification is an important component of a functional Information Retrieval ecosystem. Many current approaches to intent classification, typically framed as a classification problem, can be problematic as intents are often hard to define and thus data can be difficult and expensive to annotate. The problem is exacerbated when we need to extend the intent classification system to support multiple and in particular low-resource languages. To address this, we propose casting intent classification as a query similarity search problem -- we use previous example queries to define an intent, and a query similarity method to classify an incoming query based on the labels of its most similar queries in latent space. With the proposed approach, we are able to achieve reasonable intent classification performance for queries in low-resource languages in a zero-shot setting.
\end{abstract}




\received{7 February 2025}

\maketitle

\section{Introduction}
Intent classification of incoming queries is a crucial part of any Information Retrieval (IR) system. By detecting intent a system can empower downstream applications, such as relevance ranking and question answering, to address a user's information need to the best of their ability. However, building an intent classification system in practice can be quite challenging. For one, it is often difficult to cleanly define intents, thus making data gathering and annotation expensive. Intents can vary across different IR systems depending on the granularity of their taxonomy, and their primary use case. For example, in e-commerce search intent can be defined at the product category \& brand level \cite{qiu2022pretraining}. Such definitions of intent can frequently shift and expand as the underlying business need changes. Furthermore, real life IR systems increasingly need to support low-resource languages. In light of these challenges, we argue that it is a non-trivial task to build a functioning intent classification system that supports queries in low-resource languages, or to quickly and efficiently scale an existing one to support new languages.


In traditional supervised classification settings, one may train a classifier from the ground up or fine-tune a PLM (pre-trained language model) like BERT \cite{devlin-etal-2019-bert} with a classification output layer. While this is highly effective, it requires a non-trivial amount of high quality data which may be difficult to attain for some IR systems, especially when factoring in the potential cost to collect and/or translate these examples to support multiple different languages. 

There has been some related work to address this data scarcity issue. Many of these methods tend to take the approach of fine-tuning or pre-training a model on a high-resource area or domain, and using some sort of domain adaptation techniques or, in the case of multilingual systems, cross-lingual transfer. For example,  Yu et al. \cite{yu2021fewshot} propose a retrieval-based method in which predictions are made based on the most similar spans of text to the input query. Qiu et al.'s \cite{qiu2022pretraining} work on pre-training tasks for intent detection on the e-commerce domain also takes a similar route. These works focus on monolingual settings.

On multilingual intent classification, Khalil et al.'s \cite{khalil-etal-2019-cross} work tackles cross-lingual intent classification in a low-resource industrial setting, in which they pre-train on a high resource language (Engish) and evaluate on another high-resource language (French), or use machine translation techniques to convert the target language to the source language. This is effective but presupposes an effective translation model for the target language, which may not be available for some low-resource languages. Costello et al. \cite{costello2018multilayer} propose a technique that uses ensembling methods to improve performance on multilingual intent classification.

Finally, Bari et al. \cite{bari2021nearest} proposes a method similar to ours in which they investigate cross-lingual adaptation via nearest neighbours on a model fine-tuned on a high resource language. At inference time, they use a small set of support samples from the target language. Our work differs from theirs in that we don't fine-tune on a high-resource language and require zero annotated examples in the target low-resource language. We also evaluate in the intent classification setting including multiple low-resource languages, as opposed to the NLP/NLI setting on more popular languages on which previous works are evaluated. 

In this paper, we explore framing query intent classification as a query similarity search task. We first index queries with known intent labels by computing their latent space embeddings and indexing them. Given a new query at inference time, we then perform an approximate nearest neighbor search on the query index and resolve the new query's label using labels of its most semantically similar queries. We then experiment with the proposed approach on low-resource language queries intent classification. To summarize, our novel contributions are:

\begin{enumerate}
  \item We cast query intent classification as a query similarity search task without domain-specific fine-tuning, and analyse its strengths and drawbacks.
  \item We demonstrate the approach's feasibility in supporting intent classification across a wide spectrum of languages, including supporting low-resource languages in a zero-shot fashion, with no annotated data from the target language. 
  \item We quantify the effectiveness of the proposed approach on low-resource language query intent classification by comparing to other solutions such as machine translation and training the intent classifier directly with supervision.
\end{enumerate}

\section{Methodology}
\label{sec:method}
\subsection{Indexing Queries with Known Intent}

We begin by computing a query embedding for each query with known intent labels, and indexing them. Query embeddings can be computed  by using off-the-shelf sentence embedding models like RoBERTa \cite{DBLP:journals/corr/abs-1907-11692} for the monolingual or LaBSE \cite{DBLP:journals/corr/abs-2007-01852} for multilingual setting. To ensure a balanced query index, we maintain an equal number of queries across all intent classes. At inference time, we compute the incoming query's embedding, and fetch the top-$k$ most similar queries in the query index via approximate nearest neighbors search and use cosine similarity as the similarity measure. The final predicted label can be computed by simply taking a majority vote among labels associated with the top-$k$ most similar queries. This is formalized as the algorithm below.

\begin{algorithm}
\caption{Classification as Approximate Nearest Neighbors Search}

\begin{algorithmic}[1]

\State $\mathcal{D}_{train}, \mathcal{D}_{test}, \mathcal{M}(x), k$

\State $\mathcal{I} \gets \{(\mathcal{M}(x_r), y_r) \mid (x_r, y_r) \in \mathcal{D}_{train}\}$

\ForAll{$(x_t, y_t)\in \mathcal{D}_{test}$}

    \State $e_t \gets \mathcal{M}(x_t)$
    
    \State $(\mathcal{X}_n, \mathcal{Y}_n) \gets \text{ANN}(k, \mathcal{I}, e_t)$
    
    \State $\hat{y}_t = \text{MODE}(\mathcal{Y}_n)$

\EndFor

\end{algorithmic}
\end{algorithm}

We define $\mathcal{D}_{train}, \mathcal{D}_{test}$ as the train and test sets, respectively. $\mathcal{M}(x)$ is the encoder model output for a given input query $x$, and $\mathcal{I}$ is the generated query index.  $k$ is defined to be a custom set hyperparameter. The result $\hat{y}_t$ is the predicted label for the input query $x_t \in \mathcal{D}_{test}$.
\subsection{Label Resolution -- Picking the Optimal $k$}
\label{sec:labelres}
In order to determine an appropriate value of $k$, we run a preliminary set of experiments. Using the test set in ATIS \cite{10.3115/116580.116613}, a popular intent classification dataset, we perform a grid search across different values of $k$ from 1 - 75, and evaluate the accuracy and F1 score for each value of $k$. We repeated the experiment on 5 different pre-trained encoder models (detailed in \textit{Figure 1}) to avoid potential impacts from encoder model choices.  The results in \textit{Figure 1} show that across the 5 encoder models we tested, the average accuracy and F1 score peak around $k=31$. We keep this choice of $k$ in the following experiments. 

\begin{figure}[h]
  \includegraphics[width=0.49\linewidth]{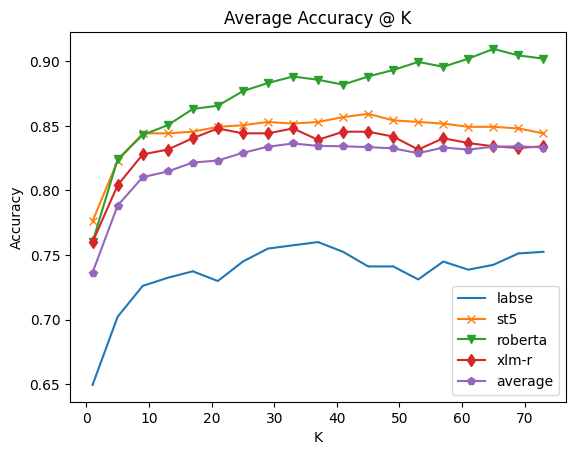}
  \includegraphics[width=0.49\linewidth]{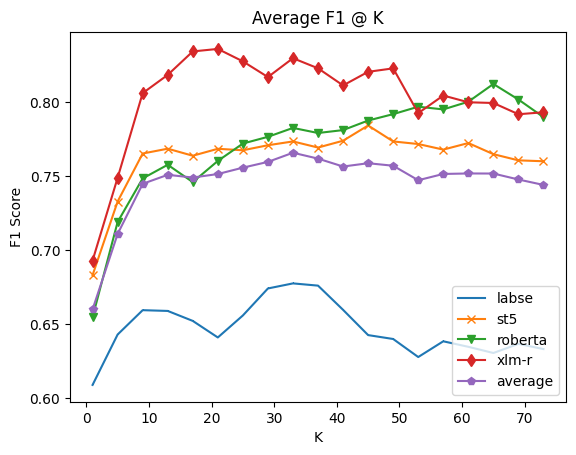}
  \caption{Average Accuracy and F1 score for values of k in the kNN majority-resolution algorithm}
  \Description{2 graphs of accuracy and F1 scores for different values of k, side by side}
\end{figure}

\section{Experiments}

\subsection{General Experiment Methodology}


For each (experiment, dataset) combination, we create a query index with FAISS \cite{douze2024faiss} following the methodology detailed in \autoref{sec:method}. We keep the number of queries indexed static by using a $\mathcal{C}-$way $\mathcal{N}$-shot approach in which for each class $\mathcal{C}$, we index $\mathcal{N}$ samples. In the multilingual setting, we extend this by indexing $\mathcal{C}\times\mathcal{N}\times\mathcal{L}$ queries, where $\mathcal{L}$ is the number of available languages. We then use the algorithm detailed in \autoref{sec:method} to compute the $\hat{y}$ for each query in the test set for the given dataset, and compute accuracy and F-1 scores by comparing the predicted $\hat{y}$ to the ground truth $y$, in a standard evaluation set up.

In order to better understand performance of the proposed approach, we also compare it with various types of alternative methods and literature results, when available. Across all experiments, we compare the proposed approach with classification head only fine-tuning. For each encoder model we test, we freeze the model weights and train a logistic regression model on top of the output embedding from the encoder. In the final set of experiments that test low-resource language query intent classification, we compare to an alternative approach that trains a classification head on a high-resource language (English) and then use an off-the-shelf machine translation model to convert test queries in the low-resource language back into English. In the following sections, we will describe the motivation and experiment-specific details for each set of experiments.



\subsection{English-only Query Intent Classification}
\label{sec:english}
\subsubsection{Motivation}
This set of experiments compares the proposed query similarity search approach with alternative methods and literature on several common English query intent classification datasets. This allows us to separate out potential impact from different language types first, and allows us to contextualise our proposed approach by comparing across a wide variety of datasets, encoder model choices, and literature results.

\subsubsection{Datasets}
We experiment on 3 common English intent classification datasets - CLINC-150 \cite{misc_clinc150_570} (excluding the Out-Of-Scope test data), HINT3 \cite{arora-etal-2020-hint3}, and ATIS \cite{10.3115/116580.116613}. We sub-sample from these datasets according to the parameters given in the relevant experimentation section.

\subsubsection{Model \& Approaches}
We test 4 different query encoders - RoBERTa-Large  \cite{DBLP:journals/corr/abs-1907-11692} (English-only), Sentence-T5 \cite{ni2021sentencet5} (English-only), XLM-RoBERTa-Large \cite{conneau2020unsupervised} (Multilingual), and LabSE \cite{DBLP:journals/corr/abs-2007-01852} (Multilingual). This gives us insights into how performance varies across monolingual v.s. multilingual models in monolingual settings. As baselines, we compare against classification heads trained on the two monolingual models, as well as literature results. 

\subsection{Multilingual Query Intent Classification}
\label{sec:multi}
\subsubsection{Motivation}

We test the proposed approach's performance in a generic multilingual setting, across a variety of different languages. Here, we assume access to a set of labelled queries across a mixture of languages, and predict the labels of test queries distributed across the same set of languages. 

\subsubsection{Datasets}
We use 2 common multilingual intent classification datasets - MultiATIS++ \cite{xu2020endtoend}, which is a translation of the ATIS dataset that covers 9 different languages, and MASSIVE \cite{fitzgerald2022massive}, which is a large-scale translation of the SLURP dataset into 50 different languages. 

\subsubsection{Model \& Approaches}
In the multilingual setting, we use the previously mentioned multilingual-aware sentence embedding models XLM-RoBERTa-Large and LabSE. As baselines, we compare against classification heads trained on the two multilingual models, as well as literature results. 

\subsection{Low-resource Language Query Intent Classification}
\subsubsection{Motivation}
\label{sec:motivation-lr}
In the final set of experiments, we test the proposed approach's performance on intent classification for queries in low-resource languages, specifically when there are little to no annotations for the target languages.

\subsubsection{Datasets}
We use the MASSIVE dataset \cite{fitzgerald2022massive}, and focus on 3 low-resource languages - Swahili (represented as sw-KE), Urdu (ur-PK), and Indonesian (id-ID). To better evaluate on the low-resource setting mentioned in \autoref{sec:motivation-lr}, we sample test queries only in the target language. We also conduct experiments in which we index no queries from the target language, thus simulating a "zero-shot" setting for that language. This is repeated for the 3 target languages.

\subsubsection{Model \& Approaches}
We test the same set of query encoder models as in \autoref{sec:multi}. However, when creating the query index, we test two different scenarios. In the first scenario, the query index is constructed with queries from all languages available in MASSIVE except for the target language. In the second scenario, we aim to better simulate the data scarcity challenge commonly seen in low-resource languages. Therefore, we construct the query index only with queries from 5 high-resource languages - English (en-EN), Chinese (zh-CN), Spanish (es-ES), French (fr-FR), and Japanese (jp-JP). To ensure query indexes in the two scenarios can be compared on an equal footing from a data perspective, we devise a specific query sampling scheme in which we control the number of queries and their semantic meaning across the two indexed sets, and vary only on the languages indexed.

As baselines, we first compare against classification heads trained on multilingual query encoder models. We also compare against an alternative translation-based approach. We use an off-the-shelf machine translation model (nllb-200-distilled-600M\cite{nllb}), translate the test queries to English, then use a trained classification head on top of an English query encoder model to make the final prediction. This is used to simulate a feasible scenario in which calling an offline machine translation model is viable, it supports the target low-resource language, and there are abundant labelled queries in a high-resource language like English.

\section{Results and Evaluation}
\subsection{English-only Query Intent Classification}
\begin{small}
\begin{table}[h]
  \begin{tabular}{cccccl}
    \toprule
    Model & Method  & ATIS & HINT3 (k=5) & CLINC-150\\
    \midrule
    XLM-R$_p$ & Sim-Search &  0.848, 0.836 & 0.470, 0.435 & 0.811, 0.808\\
    LaBSE$_p$ & Sim-Search &  0.730, 0.641 & 0.395, 0.343 & 0.829, 0.826\\
    RoBERTa$_p$ & Sim-Search & 0.866, 0.760 & 0.554, 0.509 & 0.878, 0.875\\
    RoBERTa$_p$ & Classification &  0.839, 0.703 & 0.579, 0.543 & 0.934, 0.934\\
    RoBERTa$_f$ & Classification & 0.968, 0.750 & 0.639, 0.485 & 0.953, 0.953\\
    ST5$_p$ & Sim-Search & 0.849, 0.769 & 0.600, 0.552 & 0.912, 0.909\\
    ST5$_p$ & Classification &  0.885, 0.703 & 0.584, 0.550 & 0.951, 0.950\\
    ST5$_f$ & Classification &  0.953, 0.622 & 0.460, 0.194 & 0.964, 0.964\\

    Literature & N/A & 0.989, 0.969 & 0.75, N/A  & 0.973, N/A \\
  \bottomrule
\end{tabular}
\caption{Results of English-only experiments. The reported scores are (Accuracy, F1). The $_p$/$_f$ subscripts refer to whether the training data was partial (the subset of queries indexed) or full (entire training set). Classification refers to training a classification head on the embedding for a given model. Sim-Search is our proposed approach.}
\label{tab:eval1}
\end{table}
\end{small}
In this set of experiments, unless otherwise specified, we choose $\mathcal{N}=31$ for our $\mathcal{C}-$way $\mathcal{N}$-shot sample set. The choice of $\mathcal{N}=31$ is to parallel our findings from section \ref{sec:labelres} that indexing $k=31$ was optimal for this setting. $\mathcal{C}$, the number of classes, varies on a per-dataset level. For the English-only datasets, we see promising initial results for our method. For the ATIS dataset, the off-the-shelf models perform better with our method than the classification baseline, but still falls a little short of the literature values. It is important to note that the literature models are trained on the full train set, whereas ours only uses $\sim$5\% of the dataset. We see similarly promising results for HINT3 where $k=5$. We choose to run the HINT3 experiment where $k=5$ since HINT3 has a large number of classes but a highly uneven and very sparse data distribution. Only 8/101 classes have at least 31 samples in the train set, so for a reasonable comparison we had to lower $k$. For CLINC-150, we see that the classification based approach and our approach have a significantly smaller gap. Overall, these results are promising even for English-only, but they are dataset and hyperparameter (value of $k$) dependent. To provide a more holistic view, we also compare against training the classification head on the full training dataset, as well as the best performing results available in literature - for ATIS and CLINC-150, Wang et al. \cite{wang-etal-2018-bi} and Lin et al. \cite{lin2023selective} provide strong results. Results for HINT3 were scarce, so we cite the result given in the paper \cite{arora-etal-2020-hint3}. 

\subsection{Multilingual Query Intent Classification}
\begin{small}
\begin{table}[h]
  \begin{tabular}{cccl}
    \toprule
    Model & Method & MultiATIS++ & MASSIVE\\
    \midrule
    XLM-R & Sim-Search & 0.691, 0.558 & 0.582, 0.559\\
    XLM-R$_p$ & Classification & 0.933, 0.754 & 0.729, 0.659\\
    XLM-R$_f$ & Classification & 0.982, 0.924 & 0.768, 0.726\\
    LabSE & Sim-Search & 0.669, 0.558 & 0.743, 0.721\\
    LabSE$_p$ & Classification & 0.912, 0.706 & 0.788, 0.721\\
    LabSE$_f$ & Classification & 0.975, 0.890 & 0.821, 0.787\\
    Literature & N/A & 0.958, N/A & 0.861, N/A\\
  \bottomrule
\end{tabular}
\caption{Results of multilingual experiments across various models and datasets. The reported scores are (Accuracy, F1). The $_p$ / $_f$ refer to whether the training data was partial (same as the set of queries indexed for our approach) or full (entire training set).}
\label{tab:eval2}
\end{table}
\end{small}
 In many cases, especially for the LabSE model, we see that our method is able to perform similarly to a standard classification based approach, primarily in terms of accuracy. In particular, we see strong results on MASSIVE (a more balanced dataset with more languages) with our approach -- it lags 4 points behind in accuracy for the same amount of data for the classification based approach, and less than 8 for the full dataset, which is $\sim$10x the size of the partial set. For MultiATIS++, we see mixed results, with a weaker performance relative to the classification baseline than its English counterpart, especially for XLM-RoBERTa, which is surprising, given the strong performance of our method on the English-only equivalent of this dataset. In the aggregate multilingual setting, our method generally performs slightly weaker than the classification baseline, depending on the model and dataset, and is at best on par with the baseline given the reduced data. We also, as a comparison, provide the best performing available results from the papers that proposed MultiATIS++ \cite{xu2020endtoend} and MASSIVE\cite{fitzgerald2022massive}.
 
\subsection{Low-resource Language Query Intent Classification}
\label{tab:section_lr}
\begin{small}
\begin{table}[h]
    \centering
    \begin{tabular}{cccccc}
        \toprule
        Model & Index & sw-KE & ur-PK & id-ID\\
        \midrule
        LabSE & all w/o target & 0.498, 0.483 & 0.542, 0.545 & 0.519, 0.526\\
        LabSE & high-resource & 0.486, 0.478 & 0.516, 0.518 & 0.538, 0.552\\
        XLM-RoBERTa & all w/o target & 0.072, 0.099 & 0.451, 0.430 & 0.519, 0.494\\
        XLM-RoBERTa  & high-resource & 0.119,0.125 & 0.493, 0.474 & 0.561, 0.543\\
        \bottomrule
    \end{tabular}
    \caption{Experiment results in \autoref{tab:section_lr} across different encoder models \& query indexing scenarios. Reported scores are (Accuracy, F1). Notice that classification performance on query index of high-resource language queries only is comparable to that of all but the target language.}
    \label{tab:lr1}
\end{table}
\begin{table}[h]
    \centering
    \begin{tabular}{cccccc}
        \toprule
        Model & Method & sw-KE & ur-PK & id-ID\\
        \midrule
        LabSE & Sim-Search & 0.486, 0.478 & 0.516, 0.518 & 0.538, 0.552\\
        LabSE & Classification & 0.702, 0.6127 & 0.736, 0.674 & 0.771, 0.712\\
        XLM-RoBERTa  & Sim-Search & 0.119,0.125 & 0.493, 0.474 & 0.561, 0.543\\
        XLM-RoBERTa  & Classification & 0.145, 0.114 & 0.664, 0.595	& 0.747, 0.694\\
        RoBERTa  & Translation & 0.705, 0.646 & 0.728, 0.677 & 0.736, 0.671\\
        Sentence-T5  & Translation & 0.700, 0.632 & 0.726, 0.662 & 0.729, 0.657\\
         \bottomrule
    \end{tabular}
    \caption{Experiment results in \autoref{tab:section_lr} across different encoder models \& methods. The reported scores are (Accuracy, F1). We use the high-resource query index setting to represent our proposed approach.}
    \label{tab:lr2}
\end{table}
\end{small}


Across all models and testing low-resource languages, the performance of indexing high-resource language queries only is comparable with that of indexing all available (excluding target) languages. This observation lends strong support to our hypothesis that the proposed intent classification approach can excel in a low-resource, zero-shot intent classification scenario. Under this realistic setting -- no annotated data for the target language and labels only available for high-resource language queries -- the proposed method can still perform similarly as to when annotations are available for queries in many more languages.

As indicated in \autoref{tab:lr2}, comparing to classification and translation-based approaches, the proposed query similarity search method still lags behind classification or translation-based alternatives. However, these either require model training with annotated data, or access to an external translation service that already supports the target low-resource language. In comparison, the proposed method can be rapidly deployed to solve the low-resource language use-case, without needing retraining or re-indexing of any part of the pipeline. Moreover, it can maximally reuse available data and has no translation service dependency.

\section{Conclusion and Future Work}

\subsection{Conclusion}

In this paper, we describe in detail how to construct a query index to perform intent classification via query similarity search, and conduct experiments with different off-the-shelf sentence encoder models across various datasets. Through the three sets of experiments described above, we profile the performance of the proposed approach and how it compares against alternative methods. We show that this is a practical and effective approach, particularly when being used in a zero-shot fashion in a low-resource language setting. It is straightforward and quick to implement, and requires no retraining or data annotation when on-boarding low-resource languages.  



\subsection{Future Work}

Although this is a simple yet effective approach, further exploration is still required. For example, we noted that a greater number of queries indexed per class generally leads to better outcomes. However, to quantify the nature of this relationship, as well as how it performs in situations with imbalanced class labels, is yet to be done. Moreover, we have shown the effectiveness of this method with off-the-shelf query encoder models, but are still yet to explore how fine-tuning such encoder models with available data will help.

\bibliographystyle{ACM-Reference-Format}
\bibliography{bibliography}

\end{document}